OPEN ACCESS

Asian Journal of Biological Sciences

ISSN 1996-3351
DOI: 10.3923/ajbs.2017.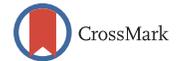

# Research Article
# Absence of Typical Haversian System from the Compact Bone of Some Reptile and Bird Species

[1]Yasser Ahmed, [2]Mohamed Abdelsaboury Khalaf and [1]Fatma Khalil

[1]Department of Histology, Faculty of Veterinary Medicine, South Valley University, Qena, Egypt
[2]Department of Anatomy and Embryology, Faculty of Veterinary Medicine, South Valley University, Qena, Egypt## Abstract

**Background and Objective:** Mammalian compact bone is composed mostly of Haversian system. Although there are many studies describing the typical Haversian system in mammals, there are few studies conducted on bones from non-mammalian species. The objective of the current study was to investigate the existence of the typical Haversian system in compact bones from reptiles and birds. **Materials and Methods:** Femora were collected from geckos, Nile monitors, sparrows, ducks and geese. Samples were then, fixed in 10% paraformaldehyde and embedded in paraffin. Paraffin sections were stained with hematoxylin and eosin or Von Kossa and then examined using light microscopy. **Results:** The current study showed different histological structures of compact bones from studied species and an absence of the typical Haversian system seen in mammals. The compact bone of geckos was mostly avascular, while Nile monitor bone was highly vascular. Sparrow bone showed one side vascular and the other side avascular. The vascularity was represented by primary vascular canals surrounded by a few irregularly arranged osteocytes inside their lacunae forming primary osteons. Bone tissue from ducks and geese were similar and were composed mostly of dense Haversian tissue, while some areas had primary osteons. **Conclusion:** The compact bone microstructure in some reptile and bird species lack the typical Haversian system described in mammals. This will be of a great importance in developing a better understanding of long bone anatomy and physiology in different species.**Key words:** Haversian system, compact bone, geckos, Nile monitors, sparrows, ducks, geese, histology

Received:                              Accepted:                              Published:

**Citation:** Yasser Ahmed, Mohamed Abdelsaboury Khalaf and Fatma Khalil, 2017. Absence of typical Haversian system from the compact bone of some reptile and bird species. Asian J. Biol. Sci., CC: CC-CC.

**Corresponding Author:** Yasser Ahmed, Department of Histology, Faculty of Veterinary Medicine, South Valley University, Qena, Egypt
Tel: +20965211223/+201098782129  Fax: +20965211223**Copyright:** © 2017 Yasser Ahmed *et al.* This is an open access article distributed under the terms of the creative commons attribution License, which permits unrestricted use, distribution and reproduction in any medium, provided the original author and source are credited.

**Competing Interest:** The authors have declared that no competing interest exists.

**Data Availability:** All relevant data are within the paper and its supporting information files.

## INTRODUCTION

The shaft of compact bone in mammals is composed of similar units known as Haversian System (HS). An early study showed that although long bones may be used for the same purposes, their Haversian system structure and arrangement are different from species to species, from bone to bone and even from side to side of the same bone[1]. The typical mammalian HS is composed of vascular canals (Haversian canals) containing blood vessels and nerves and surrounded by different types of bone lamellae. Concentric lamellae are found around the Haversian canals, periosteal lamellae are close to periosteum, endosteal lamellae are close to the endosteum and circumferential lamellae between them[2]. The Haversian canals with surrounding lamellae are commonly called Secondary Osteons (SO), which form onan earlier formed and smaller Primary Osteons (PO). The PO has no or a few lamellae smoothly merge with the surrounding bone tissue[3]. While bone lamellae of the SO are separated from surrounding bone tissue by a darkly stained cement line of bone matrix[4,5]. According to its placement, the HS is subclassified into irregular, endosteal and dense types[3]. The irregular type consists of a few SO irregularly distributed through the bone tissue. The endosteal type has SO only close to the endosteal surface, while the dense type holds numerous SO distributed in all parts of the bone[3].

The compact bone histological structure was recently used to differentiate between bone fragments from different animal species[5-8] and from human and non-human sources[9]. Although there are large numbers of histological studies of compact bone of mammals[2-13], there is very little literature on bone from reptiles and birds[14]. The current study aimed to investigate the existence of the typical HS in the compact bone of the femur from 2 non-mammalian gropus; reptiles (geckos and Nile monitors) and birds (sparrows, ducks and geese).

## MATERIALS AND METHODS

The current study started in October, 2016 in the Histology laboratory, Faculty of Veterinary Medicine, South Valley University, Egypt.

**Animals:** The current study was carried out on apparently-healthy adult yellow-bellied Egyptian house geckos (*Hemidactylus flaviviridis*), Nile monitors (*Varanus niloticus*), Egyptian house sparrow (*Passer domesticus niloticus*), Muscovy ducks (*Cairina moschata*) and Egyptian geese (*Alopochen aegyptiacus*). Geckos (n = 4) were caught alive from the office of the first author at the Faculty of Veterinary Medicine, South Valley University, Qena, Egypt. Nile monitor (n = 2) were collected from a lake in Qena City close to the Nile River. Sparrows (n = 4) were hunted with a rifle from the trees of Assiut City, Egypt. Ducks (n = 3) were obtained from a farm in Assiut city, Egypt. Geese (n = 3) were purchased from a village in Qena City, Egypt. Animals were euthanized and femora were rapidly dissected and immersed in 10% Phosphate Buffered Formalin (PBF) to avoid rapid postmortem changes. The femur was chosen because it is the longest and most loaded long bone[10].

**Sample processing:** After fixation for 3 days at 4°C, samples were decalcified in 10% (v/v) analytical grade formic acid for 1-7 days depending on the species. During this time, the specimens were tested daily for proper decalcification. After, at least, 1 day for geckos, 3 days for sparrows, 5 days for ducks and geese and 7 days for Nile monitor, bone tissue became partially decalcified and ready for sectioning as a soft tissue. Decalcified bones were crosscut at the smallest breadth of their diaphysis (midshaft) to facilitate the easier study of compact bone histology. Cross but not longitudinal sections were taken because compact bone histology is different from one side to the other side of the same bone[1].

**Histological examination:** Decalcified specimens were embedded in analytical grade paraffin and processed for histological staining according to Ahmed *et al.*[15]. Paraffin sections (5 μm-thickness) were stained with hematoxylin and eosin (H and E) as a general stain or von Kossa's method for calcium detection in bone tissue according to Bancroft and Gamble[11]. Stained sections were examined with a light microscope (Leica DMLS, Germany). Sections were photographed with Leica digital camera (Leica ICC50) using 4X, 10X, 40X and 100X objectives and pictures were captured in JPEG format.

## RESULTS

No typical Haversian system, as described in mammals by other studies Zedda *et al.*[5,12], could be seen in the compact bone from any of the examined species. Furthermore, the compact bones from the studied species were histologically different as explained below.

**Histology of the compact bone from reptiles:** Bone tissue from geckos appeared nearly avascular, except for 1 or 2 rounded vascular canals (Fig. 1a). The surrounding bone matrix contained a few number of osteocytes inside their lacunae communicating with each other through canaliculi, which appeared perpendicular to the long axis of the bone. Large areas of avascular bone tissue were characterized the gecko' compact bone. Many large deeply stained granules were seen in the bone matrix (Fig. 2b). These granules stained positive for calcium with von Kossa (Fig. 1c). One or two SO were observed in some specimens and consisted of vascular canals surrounded by 1-2 bone layers of bone lamellae containing 4-5 osteocytes embedded in lacunae and separated from the surrounding tissue by irregular cement lines (Fig. 1c). Unlike the gecko, compact bone from Nile monitors appeared highly vascular and cellular (Fig. 1d). The bone tissue contained many rounded, oval or elongated vascular canals surrounded by osteocytes with no demarcation lines separating them from the bone matrix and forming PO (Fig. 1e, f). One of the characteristic features was the presence of dense acidophilic areas around the vascular canals surrounded by a well-developed and long branched network of canaliculi (Fig. 1f). Similar to the bone matrix of geckos, many large deeply stained granules were observed and stained positive for calcium by von Kossa (Fig. 2e).

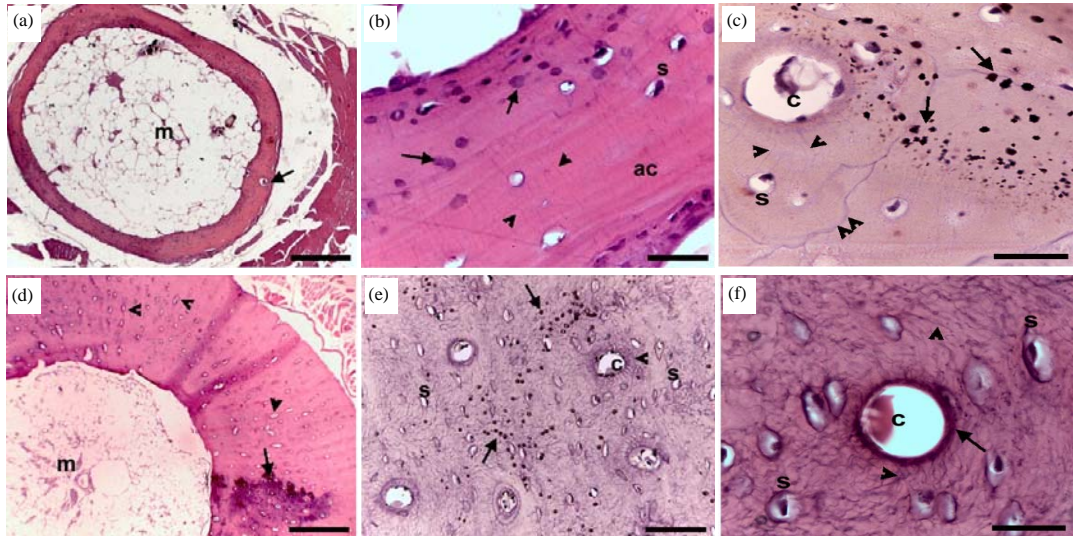

Fig. 1(a-f): Histology of the compact bone of reptilian femur, Paraffin sections from the midshaft of the femur of geckos (a-c) and Nile monitors (d-f) stained by H and E (a, b, d, f) and von Kossa (c, e), (a) Bone marrow (m) and vascular canal (arrow), (b) Osteocytes in lacunae (s), canaliculi (arrowheads), dark-stained granules (arrows) and areas of acellular bone tissue (ac), (c) Vascular canal (c), osteocytes (s), canaliculi (arrowheads), cement line (double arrowhead) of the SO and von Kossa-positive granules (arrows), (d) Bone marrow (m), vascular canals (arrowheads) and patches of dark-stained areas (arrow), (e) Vascular canal (c), canaliculi (arrowhead), osteocytes (s) and von Kossa-positive granules (arrows) and (f) Vascular canal (c), acidophilic area (arrow), canaliculi (arrowheads) and osteocytes (s)
Scale bar = 250 µm (a), 25 µm (b), 25 µm (c), 250 µm (d), 62.5 µm (e) and 25 µm (f)

**Histology of the compact bone from birds:** Compact bone from sparrows showed a small number of rounded vascular canals on one side, while the other side was a vascular (Fig. 2a). The vascular canals were surrounded by a few osteocytes merged with the surrounding tissue forming PO (Fig. 2b). Two or three small irregular SO were surrounded by 1-2 layers of bone lamellae and separated from the surrounding tissues by cement lines (Fig. 2b). Canaliculi were not clear with the light microscopy. Von Kossa-positive granules were observed in some areas of the bone tissue, especially around vascular canals (Fig. 2c). In ducks and geese the compact bone was histologically of a similar structure. It was composed mostly of dense Haversian tissue morphologically different from typical HS of mammals. The Haversian tissue in ducks and geese had a large number of small SO clustered close to each other through the bone matrix (Fig. 2d). The SO were rounded with rounded or elliptical vascular canals surrounded by 1-3 layers of bone lamellae with a small number of osteocytes inside their lacunae and were separated from bone tissue by relatively thick and darkly stained areas (Fig. 2e). These areas stained positive by von Kossa (Fig. 2f). Transfers canals were seen in some areas (Fig. 2f). Many PO were seen and some appeared to have 2 vascular canals (Fig. 2g). The PO had rounded central canals surrounded by irregularly distributed osteocytes and a network of thin canaliculi (Fig. 2h). Close to the periosteum,

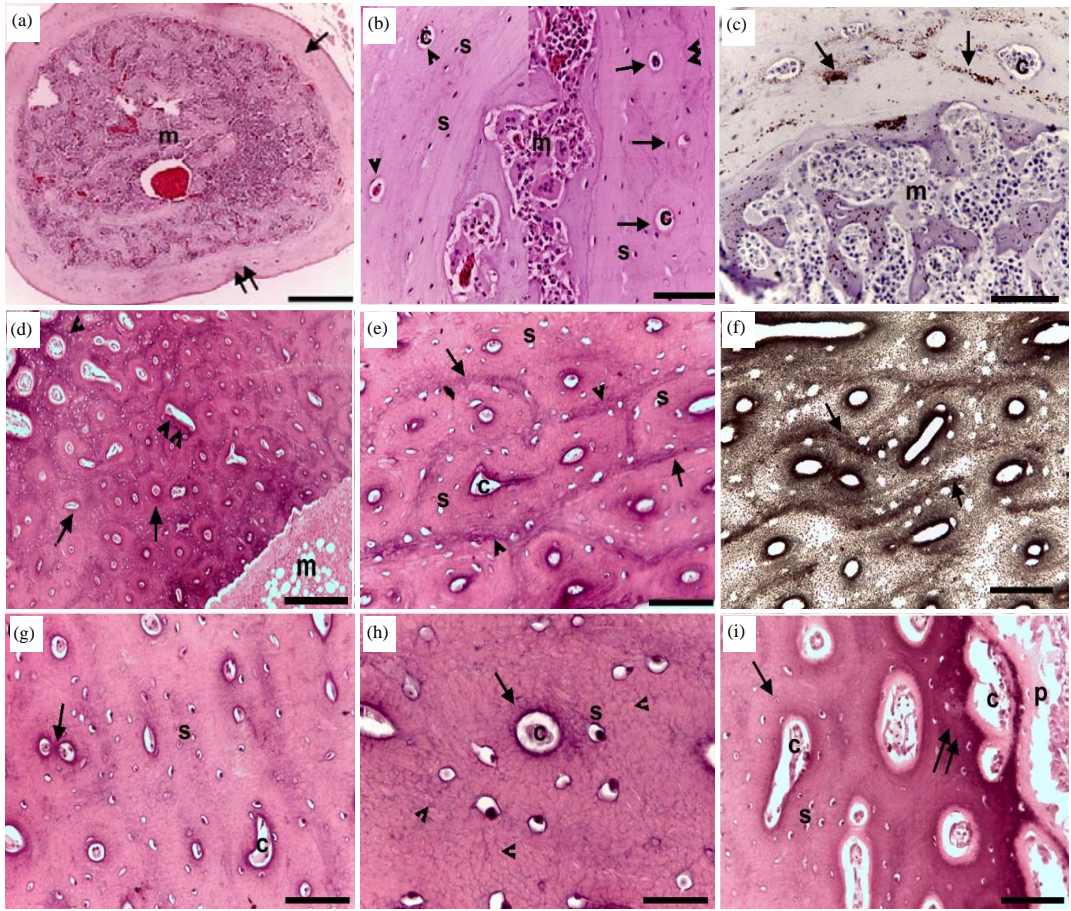

Fig. 2(a-i): Histology of the compact bone of avian femur, Paraffin sections from the midshaft of the femur of sparrows (a-c), ducks (d, g, h) and geese (e, f, i) stained by H and E (a, b, d, e, g, h, i) and von Kossa (c, f), (a) Bone marrow (m), avascular side (arrow) and vascular side (double arrows) of bone tissue, (b) Bone marrow (m), SO (arrows), PO (arrowheads), vascular canals (c), cement line (double arrowheads) and osteocytes inside lacunae (s), (c) Bone marrow (m), vascular canals (c), von Kossa-positive granules (arrows), (d) Bone marrow (m), SO (arrows), SO close to periosteum (arrowhead) and transverse canals (double arrowheads), (e) SO (arrows), cement lines (arrowheads), vascular canals (c) and osteocytes inside lacunae (s), (f) Von Kossa-positive cement line (arrows), (g) Two vascular canals (arrow) of the same PO and osteocytes (s), (h) PO (arrow), vascular canal (c), canaliculi (arrowheads) and osteocytes (s) inside lacunae and (i) SO (arrow) close to the periosteum (p), vascular canals (c), osteocytes inside lacunae (s) and dark-stained bone tissue (double arrows)

Scale bars = 250 µm (a), 62.5 µm (b), 62.5 µm (c), 250 µm (d), 62.5 µm (e), 62.5 µm (f), 250 µm (g), 25 µm (h) and 25 µm (i)

most osteons were of the secondary type and showed comparatively large and irregular vascular canals surrounded by a darkly stained bone tissue with many osteocytes (Fig. 2i).

**DISCUSSION**

In the current study, compact bone from reptilian and avian species were examined with the aim of determining the presence of typical HS as previously described in mammalian compact bone in this study. In all examined specimens, a typical HS that is found in the compact bone of mammals[5,8,12,16]. Bone from small (geckos) and large (Nile monitors) reptiles showed different histology. In geckos, the compact bone was mostly avascular except for PO. The PO appeared as 1 or 2 vascular canals surrounded by a small number of osteocytes and 1 SO could occasionally be seen in some sections. Similarly, in the femur of a Turkish lizard (*Stellagama stellio*), only 2 PO were seen and incorrectly described as SO[14]. Nile monitor bone was highly vascular and showed many rounded, oval or elongated vascular canals surrounded by randomly distributed osteocytes forming PO. A similar observation was described early in the femurs of alligators[1]. The difference in vascularity of the bone tissue from both species is related to the bone size and cortical bone growth rate[17]. The PO seen in the examined species could remain during the whole animal's life and never been replaced by De Buffrenil *et al.*[17] as the remodeling in such species is a physiologically delayed process[18]. Avian compact bone showed histological variations in sparrows versus ducks and geese. The bone tissue from the sparrow was vascular on one side and avascular on the other side. Such variation in vascularity of the same bone may be related to the mechanical stress difference applied on different sides of the bone. Only 2 or 3 small irregular SO could be observed in some sections. The low vascularity and SO appearance is related to the low-bone remodeling of birds such as sparrows. On the other hand, ducks and geese compact bone tissue was similar and consisted of dense Haversian tissue of many rounded clusters of SO separated from each other by thick dark areas. In another study, it was reported that compact bone from *Galliformes* had many small osteons with small number of bone lamellae[8]. The SO close to the periosteum had large irregular vascular canals and denser osteocytes. The PO was distributed to many areas and 2 vascular canals could be seen in the same PO. The morphological variations seen in sparrows versus ducks and geese in relation to osteonal morphology and density could be explained by the fact that these species apply biomechanical stress to the bone differently.

The histological variations seen in the compact bone of studied species may be related to the mechanical stress applied to long bones and position of the bones in the body and all could be closely correlated with the different lifestyles of these animals[10]. Even in mammals, there are variations in osteonal density according to the animal size and then mechanical load applied to bone tissue. For example, unlike large mammals, rat has very few SO and short osteonal canals[13]. Furthermore, the wild sheep long bone has a higher number and diameter of SO than domestic sheep[10].

To the best of our knowledge, this is the first histological study that provides a descriptive comparative histology of compact bone from non-mammalian species (reptiles and birds).

## CONCLUSION AND FUTURE RECOMMENDATION

It can be concluded that the typical HS described in mammalian compact bone is absent from the compact bone of some reptiles and birds. Furthermore, the histological structure of compact bone in the studied animals was widely different. Thus, the histological analysis of the long bones structure may be functionally correlated to the habits of animals, the muscular stress applied to long bones and the relative position of the bone to the animal body.

A future study will focus on the histological differences between compact bone from different animal species to investigate the possibility of using the structure of compact bone to differentiate between different species, which will be of a great importance to veterinary forensic medicine.

## SIGNIFICANCE STATEMENT

This study discovered the absence of the typical Haversian system in the compact bone of some reptilian and avian species. This is of significant importance to veterinary forensic medicine. This study will help the researchers to uncover the critical areas of long bone morphology in different species that many researchers were not previously able to explore. Thus, a new theory on long bone development may be developed.

## ACKNOWLEDGMENTS

The authors would like to thank Dr. Brad Morrison, Science Department at the Toronto District School Board, Canada and Prof. Antar Abdallah, Professor of English education at Taibah University, Saudi Arabia, for editing of the manuscript. As well, thanks Dr Fatma Abass and Mrs. Fatma Abdel Hakim, Department of Anatomy, Faculty of Veterinary Medicine, South Valley University, Qena, Egypt for kindly providing the goose and duck specimens.